\pdfoutput=1
\documentclass[]{osa-article}

\journal{osajournal}
\articletype{Research Article}
\usepackage{empheq}
\usepackage{bm}
\usepackage[normalem]{ulem}
\usepackage{booktabs}
\usepackage{upgreek}
\usepackage[mathscr]{euscript}

\let\originalleft\left
\let\originalright\right
\renewcommand{\left}{\mathopen{}\mathclose\bgroup\originalleft}
\renewcommand{\right}{\aftergroup\egroup\originalright}
\DeclareMathAlphabet{\mathcal}{OMS}{cmsy}{m}{n}

\providecommand{\mb}[1]{\mathbf{#1}}

\providecommand{\mc}[1]{\mathcal{#1}}
\providecommand{\ro}{\boldsymbol{\mathfrak{r}}_o}

\newcommand{\mypar}{\parallel}
\providecommand{\ropar}{r_o^{\mypar}}
\providecommand{\roperp}{\mathbf{r}_o^{\bot}}

\providecommand{\so}{\mathbf{\hat{s}}_o}

\providecommand{\rbm}[1]{r_b^{\text{m}}}
\providecommand{\rd}{\mathbf{r}^{\bot}_d}

\providecommand{\mh}[1]{\mathbf{\hat{#1}}}
\providecommand{\mbb}[1]{\mathbb{#1}}
\providecommand{\bs}[1]{\boldsymbol{#1}}

\providecommand{\lmsum}{\sum_{\ell=0}^\infty\sum_{m=-\ell}^{\ell}}

\providecommand{\sothree}{\mbb{SO}(3)}
\providecommand{\smol}{\mc{D}_{\mb{v}}}

\begin{document}

\title{Spatio-angular fluorescence microscopy\\ III. Constrained angular diffusion, polarized excitation, and high-NA imaging}

\author{Talon Chandler,\authormark{1,*} Hari Shroff,\authormark{2,3} Rudolf Oldenbourg,\authormark{3} and Patrick La Rivi\`ere\authormark{1,3}}

\address{\authormark{1}University of Chicago, Department of Radiology, Chicago, Illinois 60637, USA\\
  \authormark{2}Section on High Resolution Optical Imaging, National Institute of Biomedical Imaging and Bioengineering, National Institutes of Health, Bethesda, Maryland 20892, USA\\
  \authormark{3}Marine Biological Laboratory, Bell Center, Woods Hole, Massachusetts 02543, USA}

\email{\authormark{*}talonchandler@talonchandler.com} %% email address is required

\begin{abstract*}
  We investigate rotational diffusion of fluorescent molecules in angular
  potential wells, the excitation and subsequent emissions from these diffusing
  molecules, and the imaging of these emissions with high-NA aplanatic optical
  microscopes. Although dipole emissions only transmit six low-frequency angular
  components, we show that angular structured illumination can alias higher
  frequency angular components into the passband of the imaging system. We show
  that the number of measurable angular components is limited by the
  relationships between three time scales: the rotational diffusion time, the
  fluorescence decay time, and the acquisition time. We demonstrate our model by
  simulating a numerical phantom in the limits of fast angular diffusion, slow
  angular diffusion, and weak potentials.
\end{abstract*}

\section{Introduction}
Rotational diffusion plays an important role in every fluorescence experiment.
Stokes' 1852 investigation of fluorescence (which led him to coin the word
``fluorescence'') reported no apparent polarization of the light emitted by a
fluorescing solution of quinine \cite{stokes1852}. We now understand that his
observation reflects the relative time scales of angular diffusion, fluorescence
decay, and measurement acquisition \cite{jameson2010}. Angular diffusion of
quinine ($\sim$0.3 ns rotational relaxation time) is fast compared to its
fluorescence lifetime ($\sim$20 ns), which is fast compared to Stokes'
acquisition time ($\sim$0.1 s for human vision). Even though each individual
emission is polarized, diffusive reorientation of each fluorophore results in
randomly polarized emissions that result in no apparent polarization when
averaged over the measurement time.

These relationships were elucidated by several investigators in the 1920s.
Weigert demonstrated that decreasing the rotational mobility of fluorescent
molecules (by increasing the viscosity of the solvent or decreasing the
temperature) resulted in increasingly polarized fluorescent emissions
\cite{weigert1920}. Wawilow and Lewschin observed that different dyes displayed
varying relationships between the rotational mobility and the polarization of
the fluorescent emissions \cite{wawilow1923}, and Francis Perrin explained these
variations by accounting for the fluorescence lifetime of the fluorophores
\cite{perrin1926}. Perrin's synthesis inspired Weber to develop modern
fluorescence polarization assays for biological applications \cite{weber1952a,
  weber1952b}. See Jameson's review \cite{jameson2010} for English summaries of
the papers cited in this paragraph.

Since Weber's work, fluorescence polarization assays have been used to deduce
information from a wide range of samples in solution---see Lakowicz
\cite[chapters 10-12]{lakowicz2006} for a review. More recently, fluorescence
polarization imaging assays have been developed to image rotationally
constrained fluorophores that label biological structures \cite{axelrod1979,
  dale1999, siegel2003, demay2011a, mattheyses2010, brasselet2011,
  zhanghao2019}. Furthermore, breakthroughs in single-molecule localization
microscopy have led to assays that measure the position, orientation, and
rotational dynamics of single molecules \cite{ha1996, forkey2003, toprak2006,
  aguet2009a, zhang2018a, backer2019}. All of these techniques use a
model of rotational diffusion and the imaging process to interpret the collected
data, and any mismatch between the model and the experiment 
could limit the accuracy of these interpretations.

Several recent works have modeled the images created by rotating single
molecules under angular constraints \cite{lew2013, backer2015, stallinga2015,
  mehta2016}, and this paper refines and extends these models. First, we
consider angular potentials more general than those that are rotationally
symmetric about a single axis. Modeling general potentials reduces the number of
assumptions required to interpret data and creates opportunities for
designing instruments that can draw new conclusions. Second, we consider in
detail how the angular potential affects angular diffusion. Existing works have
assumed that angular diffusion can be described by a monoexponential decay,
while here we use the Smoluchowski equation to show that angular diffusion is
multi-exponential with time constants that depend on the potential. Third, we
consider the effects of fluorescence saturation on the spatio-angular imaging
process. We show that exploiting saturation can enable measurements of
high-frequency angular components. Finally, we efficiently model arbitrary
spatio-angular distributions of fluorescent emitters including but not limited
to single molecules. These modeling improvements make several new predictions
that may guide future experiments and improve the interpretation of existing
data.

In the previous two papers of this series \cite{chandler2019a, chandler2019b}, we
described the organization of our theory, described spatio-angular imaging
operators and how they can be expressed in different bases, then calculated
spatio-angular imaging operators for a paraxial 4$f$ imaging system. In this
paper, we build on our framework and incorporate angular diffusion within a potential, polarized excitation, and high-NA imaging. 

The paper is organized as follows. In Section \ref{sec:theory} we develop models
for spatio-angular diffusion, excitation, emission, and imaging. After introducing our
notation (Section \ref{sec:notation}), we build on the work of Jones
\cite{jones2003a} and Schulten et al. \cite{schulten2000} to describe angular
diffusion of one-state dipoles within asymmetric (Section \ref{sec:arbitrary})
and symmetric potentials (Section \ref{sec:symmetric}). Next, we describe the
diffusion of two-state molecules and their emissions under strong (Section
\ref{sec:two-state}), and weak excitation (Section
\ref{sec:weak}--\ref{sec:lastangle}). In Section \ref{sec:results} we create a
numerical phantom, specify an imaging system, then simulate imaging results.
Finally, in Section \ref{sec:discussion} we discuss our results and their
implications.

\section{Theory}\label{sec:theory}
\subsection{Notation}\label{sec:notation}
We use roman fonts for scalars and functions (e.g. $t,f$), bold fonts for
vectors (e.g. $\mb{f},\mb{s}$), and blackboard bold for manifolds and vector
spaces (e.g. $\mbb{S}^2$, $\mbb{R}^3$). We use hats to denote unit vectors (e.g.
$\mh{s}$, $\mh{e}$), and we use $\{\mh{e}_i\}$ to denote a set of orthonormal
standard basis vectors.

We briefly review notation for functions that map points on the sphere
$\mbb{S}^2$ onto the real numbers $\mbb{R}$. We denote these spherical functions
by $f(\mh{s})$ where $\mh{s} \in \mbb{S}^2$, and we denote their associated
Hilbert-space vectors by $\mb{f} \in \mbb{L}_2(\mbb{S}^2)$. We define an inner
product for this Hilbert space as
\begin{align}
  (\mb{f}_1, \mb{f}_2) = \int_{\mbb{S}^2}d\mh{s}\,f_1(\mh{s})f_2(\mh{s}),
\end{align}
and we use this inner product to confirm that the non-denumerable set of standard basis vectors $\{\mh{e}(\mh{s})\} = \{[1,0,0\ldots], [0,1,0,\ldots],\ldots,[\ldots,0,0,1]\}$ satisfy
\begin{align}
  \left(\mh{e}(\mh{s}), \mh{e}(\mh{s}')\right) = \delta(\mh{s} - \mh{s}'),
\end{align}
where $\delta(\mh{s} - \mh{s}')$ is the Dirac delta on the sphere. We can construct an alternative orthonormal basis using the real-valued spherical harmonic functions $Y_{\ell m}(\mh{s})$ which satisfy
\begin{align}
  \int_{\mbb{S}^2}d\mh{s}\,Y_{\ell m}(\mh{s})Y_{\ell' m'}(\mh{s}) = \delta_{\ell\ell'}\delta_{mm'},
\end{align}
where $\ell \in \{0,1,2,\ldots\}$ and
$m \in \{-\ell, -\ell + 1, \ldots \ell - 1, \ell\}$. The new basis vectors are
\begin{align}
  \mh{E}_{\ell m} = \int_{\mbb{S}^2}d\mh{s}\, Y_{\ell m}(\mh{s})\mh{e}(\mh{s}),
\end{align}
which satisfy
\begin{align}
  \left(\mh{E}_{\ell m}, \mh{E}_{\ell'm'}\right) &= \delta_{\ell\ell'}\delta_{mm'},\\
  \left(\mh{e}(\mh{s}), \mh{E}_{\ell m}\right) = \left(\mh{E}_{\ell m}, \mh{e}(\mh{s}) \right) &= Y_{\ell m}(\mh{s}).
\end{align}
We can expand arbitrary Hilbert-space vectors $\mb{f} \in \mbb{L}_2(\mbb{S}^2)$
in either basis as
\begin{align}
  \mb{f} &= \int_{\mbb{S}^2}d\mh{s}\,f(\mh{s})\mh{e}(\mh{s}) = \sum_{\ell = 0}^{\infty}\sum_{m=-\ell}^{\ell}F_{\ell m}\mh{E}_{\ell m}.\label{eq:expsh}
\end{align}
The coefficients $f(\mh{s})$ can be found by taking the inner product of both sides of Eq. \eqref{eq:expsh} with the basis vectors $\mh{e}(\mh{s})$ and exploiting orthonormality
\begin{align}
  f(\mh{s}) = (\mh{e}(\mh{s}), \mb{f}).
\end{align}
We can proceed similarly for the coefficients $F_{\ell m}$ then write these coefficients in terms of $f(\mh{s})$
\begin{align}
  F_{\ell m} &= (\mh{E}_{\ell m},\mb{f}) = \int_{\mbb{S}^2}d\mh{s}\,(\mh{E}_{\ell m},\mh{e}(\mh{s}))(\mh{e}(\mh{s}),\mb{f}) =  \int_{\mbb{S}^2}d\mh{s}\,Y_{\ell m}(\mh{s})f(\mh{s}), 
\end{align}
which is usually called the spherical Fourier transform.

% We will also consider square-integrable function on the plane
% $g(\mb{r}) \in \mbb{L}_2(\mbb{R}^2)$ where $\mb{r} \in \mbb{R}^2$ is a
% two-dimensional vector. We define an inner product
% \begin{align}
%   (\mb{g}_1, \mb{g}_2) = \int_{\mbb{R}^2}d\mb{r}\, g_1^*(\mb{r})g_2(\mb{r}),
% \end{align}
% and use it to confirm that the non-denumerable set of standard basis vectors $\{\mh{e}(\mb{r})\} = \{[1,0,0\ldots], [0,1,0,\ldots],\ldots,[\ldots,0,0,1]\}$ satisfy
% \begin{align}
%   (\mh{e}(\mb{r}), \mh{e}(\mb{r}')) = \delta(\mb{r} - \mb{r}').
% \end{align}
% We can construct an alternative orthonormal basis using the complex exponentials which satisfy
% \begin{align}
%   \int_{\mbb{R}^2} d\mb{r}\,\exp[-i2\pi \mb{r}\cdot\bs{\nu}]\exp[i2\pi \mb{r}\cdot\bs{\nu}'] = \delta(\bs{\nu} - \bs{\nu}'),
% \end{align}
% where $\bs{\nu} \in \mbb{R}^2$. The new basis vectors are
% \begin{align}
%   \mh{E}(\bs{\nu}) = \int_{\mbb{R}^2}d\mb{r}\,\exp[i2\pi\mb{r}\cdot\bs{\nu}]\mh{e}(\mb{r}), 
% \end{align}
% which satisfy
% \begin{align}
%   (\mh{E}(\bs{\nu}), \mh{E}(\bs{\nu}')) &= \delta(\bs{\nu} - \bs{\nu}'), \\
%   (\mh{e}(\mb{r}), \mh{E}(\bs{\nu})) &= \exp[i2\pi \mb{r}\cdot \bs{\nu}].  
% \end{align}

We denote Hilbert-space operators with capital calligraphic letters $\mc{H}$.
Hilbert-space operators act on Hilbert-space vectors to create other
Hilbert-space vectors $\mb{g} = \mc{H}\mb{f}$, and we can express the action of
$\mc{H}$ concretely by choosing a basis for $\mb{g}$ and $\mb{f}$. For example,
if $\mc{H}:\mbb{L}_2(\mbb{S}^2)\rightarrow\mbb{L}_2(\mbb{R}^2)$ then we can choose the standard basis for both the input and output spaces and write
\begin{align}
  \mb{g} &= \mc{H}\mb{f},\\
  (\mh{e}(\mb{r}),\mb{g}) &= \int_{\mbb{S}^2}d\mh{s}\,(\mh{e}(\mb{r}),\mc{H}\mh{e}(\mh{s}))(\mh{e}(\mh{s}), \mb{f}),\\
  g(\mb{r}) &= \int_{\mbb{S}^2}d\mh{s}\,h(\mb{r},\mh{s})f(\mh{s}),
\end{align}
where $h(\mb{r},\mh{s}) = (\mh{e}(\mb{r}),\mc{H}\mh{e}(\mh{s}))$ are the \textit{standard entries} of $\mc{H}$.

We can calculate the entries of $\mc{H}$ in a different basis by relating them to the standard entries. For example
\begin{align}
H_{\ell m}(\mb{r}) \equiv (\mh{e}(\mb{r}), \mc{H}\mh{E}_{\ell m}) = \int_{\mbb{S}^2}d\mh{s}\, \left(\mh{e}(\mb{r}), \mc{H}\mh{e}(\mh{s})\right)(\mh{e}(\mh{s}), \mh{E}_{\ell m}) = \int_{\mbb{S}^2}d\mh{s}\, h(\mb{r},\mh{s}')Y_{\ell m}(\mh{s}).
\end{align}
Finally, we denote adjoint operators with a dagger $\dagger$ using the definition 
\begin{align}
  (\mb{f}_1, \mc{H}\mb{f}_2) = (\mc{H}^\dagger\mb{f}_1, \mb{f}_2).
\end{align}

\subsection{Dipole angular diffusion in arbitrary potentials}\label{sec:arbitrary}
Consider a rigid molecule with orientation $\mb{R} \in \sothree$---a $3\times 3$
orthogonal matrix with determinant $+1$. We let $\omega(\mb{R}, t)$ denote the
probability of finding the molecule in orientation $\mb{R}$ at time $t$. Our
goal is to develop a useful model for the time evolution of $\omega(\mb{R}, t)$
given an initial condition $\omega(\mb{R}, 0)$. 

We start by assuming that the molecule's orientation can be completely described
by a single absorption/emission dipole axis $\mh{s}$. To apply this assumption,
we parameterize the molecule's orientation using an axis-angle representation
$\mb{R} = (\mh{s}, \psi)$, where $\mh{s} \in \mbb{S}^2$ specifies the dipole
axis and $\psi \in [0,2\pi)$ specifies a rotation about $\mh{s}$. With this
parameterization we can apply the assumption by ignoring $\psi$ and considering
the time evolution of $\omega(\mh{s},t)$. We also know that dipole
absorber/emitters are symmetric under inversion, so we constrain our
distributions to be of the form $\omega(\mh{s},t) = \omega(-\mh{s},t)$. Finally,
we normalize our probability distribution using
\begin{align}
  \int_{\mbb{S}^2}d\mh{s}\,\omega(\mh{s},t) = 1.
\end{align}

We can model the time evolution of $\omega(\mh{s}, t)$ using the \textit{Smoluchowski equation}
\begin{align}
    \frac{\partial \omega(\mh{s}, t)}{\partial t} &=  \nabla \cdot \textbf{\textit{D}}(\mh{s})\left[\nabla \omega(\mh{s}, t) + \beta \omega(\mh{s},t)\nabla v(\mh{s})\right],\label{eq:smol}
\end{align}
where $v(\mh{s})$ is an arbitrary angular potential (for example, from an
external field or a molecular bond), $\nabla$ is the spherical gradient
operator, $\nabla\cdot$ is the spherical divergence operator,
$\textbf{\textit{D}}(\mh{s})$ is an orientation-dependent angular diffusion
tensor, and $\beta = 1/k_BT$ with $k_B$ Boltzmann's constant and $T$
temperature. Although we point readers elsewhere for a derivation
\cite{risken1996, schulten2000, coffey2012}, Eq. \eqref{eq:smol} is plausible at
a glance. The first term in brackets models diffusion down a concentration
gradient, and the second term models torques due to the potential. The
orientation-dependent angular diffusion tensor scales and rotates the gradients.
Next, the divergence sums over all neighboring orientations to find the total
change in $\omega(\mh{s}, t)$. We note that Eq. \eqref{eq:smol} assumes that
inertial terms are negligible and that the torques can be related to a scalar
potential.

Next, we assume that the molecule behaves like a \textit{spherical rotor}---the
diffusion tensor is homogeneous (independent of $\mh{s}$) and isotropic
(independent of angular diffusion direction)---so we can replace $\textbf{\textit{D}}(\mh{s})$ with a single constant $D$
\begin{align}
    \frac{\partial \omega(\mh{s}, t)}{\partial t} &= D \nabla \cdot \left[\nabla \omega(\mh{s}, t) + \beta \omega(\mh{s},t)\nabla v(\mh{s})\right].\label{eq:smol2}
\end{align}
This assumption is widely used in fluorescence microscopy \cite{lew2013,
  backer2015, stallinga2015, mehta2016}, and it provides a reasonable
approximation for globular emitters like green fluorescent protein. Other
investigators have modeled fluorescence from non-spherical rotors in solution
\cite{loman2010, lakowicz2006} and non-fluorescent diffusion of non-spherical
rotors in a potential \cite{coffey2012}, while here we focus on modeling
fluorescence from spherical rotors in a potential.

We can rewrite Eq. \eqref{eq:smol2} in terms of Hilbert-space vectors and operators by collecting the $\mh{s}$ dependence of $\omega(\mh{s},t)$ and $v(\mh{s})$ into boldface vectors $\bs{\omega}(t) = \int_{\mbb{S}^2}d\mh{s}\,\omega(\mh{s},t)\mh{e}(\mh{s})$ and $\mb{v} = \int_{\mbb{S}^2}d\mh{s}\,v(\mh{s})\mh{e}(\mh{s})$ then writing
\begin{align}
  \frac{\partial \bs{\omega}(t)}{\partial t} &= D \nabla \cdot \left[\nabla \bs{\omega}(t) + \beta\bs{\omega}(t)\nabla\mb{v}\right] = -\smol\bs{\omega}(t),\label{eq:smol3} 
\end{align}
where $\smol = -D \nabla \cdot \left[\nabla + \beta\nabla\mb{v}\right]$ is the Smoluchowski operator with arbitrary potential $\mb{v}$ and
the negative sign is included for convenience.

Eq. \eqref{eq:smol3} is a homogeneous system of linear first-order differential
equations. A typical approach is to expand $\bs{\omega}$ into a linear combination of
eigenfunctions of $\smol$, but it is not obvious that a complete set of
eigenfunctions exists. To show that a complete set of eigenfunctions does exist
we follow Schulten et al. \cite{schulten2000} and rewrite $\smol$ as
\begin{align}
  \smol = -D\nabla\cdot\exp(-\beta\mb{v})\nabla\exp(\beta\mb{v}). 
\end{align}
In this form it is straightforward to confirm that 
\begin{align}
  \mc{W}\smol\mc{W}^{-1} = \mc{B}^{\dagger}\mc{B},\label{eq:simpos}
\end{align}
where
\begin{align}
  \mc{W} &= \exp(\beta\mb{v}/2),\\
  \mc{B} &= \sqrt{D}\exp(-\beta\mb{v}/2)\nabla\exp(\beta\mb{v}/2),\\
  \mc{B}^{\dagger} &= -\sqrt{D}\exp(\beta\mb{v}/2)\nabla\cdot\exp(-\beta\mb{v}/2),
\end{align}
and we have used the following operator identity (the adjoint of the gradient is the negative divergence)
\begin{align}
 \nabla^\dagger = -\nabla\cdot. 
\end{align}
Eq. \eqref{eq:simpos} shows that $\smol$ is similar to a Hermitian operator
\cite[ch.~1.4]{barrett2004}, and Hermitian operators have real non-negative
eigenvalues $\lambda_{\mb{v},i}$ and a complete set of orthogonal eigenfunctions
$\bs{\psi}_{\mb{v},i}$ that satisfy
\begin{align}
  \mc{W}\smol\mc{W}^{-1}\bs{\psi}_{\mb{v},i} = \lambda_{\mb{v},i}\bs{\psi}_{\mb{v},i}.
\end{align}
Applying $\mc{W}^{-1}$ to both sides yields
\begin{align}
  \smol\mc{W}^{-1}\bs{\psi}_{\mb{v},i} &= \lambda_{\mb{v},i}\mc{W}^{-1}\bs{\psi}_{\mb{v},i},\\
  \smol\bs{\phi}_{\mb{v},i} &= \lambda_{\mb{v},i}\bs{\phi}_{\mb{v},i},
\end{align}
which shows that we can find a complete (though not necessarily orthogonal) set of
eigenfunctions for $\smol$ by calculating
$\bs{\phi}_{\mb{v},i} = \mc{W}^{-1}\bs{\psi}_{\mb{v},i}$. Additionally, we have
shown that $\smol$ has real non-negative eigenvalues, so we can labels its eigenvalues in order
\begin{align}
  0 \leq \lambda_{\mb{v},0} \leq \lambda_{\mb{v},1} \leq \lambda_{\mb{v},2} \leq \lambda_{\mb{v},3} \leq \cdots.
\end{align}

Now that we have confirmed that $\smol$ has a complete set of eigenvectors, we can write the general solution of Eq. \eqref{eq:smol3} as 
\begin{align}
  \bs{\omega}(t) = \sum_{i=0}^\infty c_{\mb{v},i}\bs{\phi}_{\mb{v},i}\exp(-\lambda_{\mb{v},i} t),\label{eq:solution}
\end{align}
where $c_{\mb{v},i}$ are constants determined by the initial condition.

From statistical mechanics we expect the Boltzmann distribution to be a
steady-state solution. We can confirm this expectation by plugging the
Boltzmann distribution
\begin{align}
  \bs{\phi}_{\mb{v},0} &= Z_{\mb{v}}^{-1}\exp(-\beta\mb{v}),
\end{align}
where $Z_{\mb{v}}$ is the partition function
\begin{align}
  Z_{\mb{v}} = \int_{\mbb{S}^2}d\mh{s}\,\exp(-\beta v(\mh{s})),
\end{align}
into Eq. \eqref{eq:smol3} and confirming that it is an eigenfunction of $\smol$
with a zero eigenvalue. We also expect the Boltzmann distribution to be the
unique steady-state solution---the only eigenfunction with a zero eigenvalue. We
point readers elsewhere for a physical argument that this is true
\cite{schulten2000}, but we remark that a single steady-state solution depends
on $v(\mh{s})$ being square-integrable. For example, a non-square-integrable
potential could have two potential wells separated by an infinite potential, and
in this case we would expect multiple steady-state solutions.

Finally, we calculate the coefficients $c_{\mb{v},i}$ in terms of the initial condition $\bs{\omega}(0)$. The naive approach of taking the inner product of both sides of Eq.  \eqref{eq:solution} with the eigenfunctions $(\bs{\phi}_{\mb{v},i},\cdot)$ will fail because the eigenfunctions are not orthogonal. Instead, we construct a biorthogonal set by solving the eigenvalue problem for the adjoint Smoluchowski operator. If we write the adjoint Smoluchowski operator in the form
\begin{align}
  \smol^{\dagger} = -D\exp(\beta\mb{v})\nabla\cdot\exp(-\beta\mb{v})\nabla,
\end{align}
then it is straightforward to confirm that 
\begin{align}
  \smol^{\dagger}\left(\frac{\bs{\phi}_{\mb{v},i}}{\bs{\phi}_{\mb{v},0}}\right) = \lambda_i\left(\frac{\bs{\phi}_{\mb{v},i}}{\bs{\phi}_{\mb{v},0}}\right),
\end{align}
where the division of Hilbert-space vectors is elementwise. Therefore,
$(\bs{\phi}_{\mb{v},i}/\bs{\phi}_{\mb{v},0})$ are eigenfunctions of
$\smol^{\dagger}$, and these functions form a biorthogonal set with the
eigenfunctions of $\smol$. We can normalize so that these functions form a
biorthonormal set that satisfies
\begin{align}
  \left(\frac{\bs{\phi}_{\mb{v},i}}{\bs{\phi}_{\mb{v},0}},\bs{\phi}_{\mb{v},j}\right) = \delta_{ij}.\label{eq:biorthro}
\end{align}
Now we can take the inner product of both sides of Eq. \eqref{eq:solution} with the eigenfunctions of $\smol^{\dagger}$ and solve for $c_{\mb{v},i}$ in terms of the initial condition
\begin{align}
  c_{\mb{v},i} = \left(\frac{\bs{\phi}_{\mb{v},i}}{\bs{\phi}_{\mb{v},0}},\bs{\omega}(0)\right).
\end{align}
Therefore, the solution takes the form 
\begin{align}
  \bs{\omega}(t) = \sum_{i=0}^\infty \left(\frac{\bs{\phi}_{\mb{v},i}}{\bs{\phi}_{\mb{v},0}},\bs{\omega}(0)\right)\bs{\phi}_{\mb{v},i}\exp(-\lambda_{\mb{v},i} t).\label{eq:solution2}
\end{align}

We can use these results to draw several conclusions for diffusion under
arbitrary potentials. The fact that the eigenvalues are real implies that the
solutions will never oscillate---this is expected since we are ignoring inertial
effects. The fact that the eigenvalues are positive except for a single zero
eigenvalue implies that our solutions will always decay to the Boltzmann
distribution $\bs{\omega}(t\rightarrow\infty) = \bs{\phi}_{\mb{v},0}$. The smallest
non-zero eigenvalue $\lambda_{\mb{v},1}$ will set the time scale of the decay,
so we know that $\bs{\omega}(t \gg 1/\lambda_{\mb{v},1}) \approx \bs{\phi}_{\mb{v},0}$.
Finally, the decay of $\bs{\omega}(t)$ will be monoexponential if the initial
condition is either an eigenfunction of $\smol$ or a linear combination of
eigenfunctions of $\smol$ that share an eigenvalue.

\subsection{Dipole angular diffusion in symmetric potentials}\label{sec:symmetric}
Next, we consider how symmetric potentials can constrain the form of the
solution. If the potential is rotationally symmetric ($v(\mh{s})$ is constant)
then the Smulochowski equation reduces to the diffusion equation
\begin{align}
    \frac{\partial \bs{\omega}(t)}{\partial t} &= D\Delta\bs{\omega}(t) = -\mc{D}_c\bs{\omega}(t),\label{eq:smol4}
\end{align}
where $\Delta$ is the spherical Laplacian, and $\mc{D}_c = -D\Delta$. This
equation has a well-known solution---the eigenfunctions of $\mc{D}_c$ that
satisfy the biorthonormality relation Eq. \eqref{eq:biorthro} are the
renormalized spherical harmonics
$\bs{\phi}_{c,\ell m} = \mh{E}_{\ell m}/\sqrt{4\pi}$ with eigenvalues
$\lambda_{c,\ell m} = D\ell(\ell+1)$, which we can plug into Eq.
\eqref{eq:solution2} to find that
\begin{align}
  \bs{\omega}(t) &= \sum_{\ell=0,2,4\ldots}^\infty\sum_{m=-\ell}^\ell\left(\mh{E}_{\ell m}, \bs{\omega}(0)\right)\mh{E}_{\ell m}\exp(-D(\ell(\ell+1) t)).\label{eq:dcsol}
\end{align}
Eq. \eqref{eq:dcsol} has a simple form when expressed in terms of the spherical harmonic coefficients $\Omega_{\ell m}$
\begin{align}
  (\mh{E}_{\ell m},\bs{\omega}(t)) \equiv \Omega_{\ell m}(t) &= \Omega_{\ell m}(0)\exp(-D\ell(\ell+1)t). \label{eq:solutionc}
\end{align}

An essential feature of this solution is that each eigenvalue
$\lambda_{c,\ell m}$ forms a multiplet with $2\ell + 1$ other
eigenvalues indexed by $m$. This fact allows us to split the single eigenvalue
index $i$ in Eq. \eqref{eq:solution2} into a pair of indices $(\ell, m)$ in Eq.
\eqref{eq:dcsol}.

The multiplicity of eigenvalues reduces the number of decay components in the
solution. For example, if the initial condition is bandlimited to $\ell = 2$,
that is ($\mh{E}_{\ell m}, \bs{\omega}(0)) \neq 0$ for $\ell = 0$ and $\ell=2$ only,
then the 6-dimensional initial distribution will decay towards the Boltzmann
distribution monoexponentially with time constant $(6D)^{-1}$. 

We can predict eigenvalue multiplets by studying the \textit{symmetry group of
  the operator} $\smol$\cite[ch.~6.7]{barrett2004}, and we will use the
rotationally symmetric example $\mc{D}_c$ to illustrate this process. First, we
identify the symmetry group of the operator by finding the set of operators that
commute with $\smol$. All three-dimensional rotation operators commute with
$\mc{D}_c$ because rotating the potential leaves it unchanged. Therefore, we
identify the symmetry group of $\mc{D}_c$ as $\sothree$. Next, we find the
\textit{irreducible representations} of the symmetry group---sets of irreducible
matrices assigned to each group element where matrix multiplication reproduces the group composition rule. An irreducible representation that uses
$N\times N$ matrices is said to be $N$-dimensional. Irreducible representations
can be calculated from scratch, but in practice they can almost always be found
in the literature \cite{tung1985, hamermesh1989}. The irreducible
representations of $\sothree$ are the Wigner D-matrices
$\textbf{D}_{\ell}(\mb{R})$ where $\ell = 0, 1/2, 1, 3/2, 2, \ldots$ indexes the
$(2\ell+1)$-dimensional irreducible representations and $\mb{R}$ indexes the
elements of $\sothree$. Finally, eigenvalue $N$-plets correspond to the
$N$-dimensional irreducible representations of the symmetry group of $\smol$.
$\mc{D}_c$ has irreducible representations of integer dimension, so there
will be at most an $N$-plet for each natural number $N$.

Some of the multiplets may not appear in the solution due to
symmetries of the distribution $\omega(\mb{R})$. For example, we expect
$\omega(\mh{s},\psi) = \omega(\mh{s},\psi + 2\pi)$ which implies that
$(\bs{\phi}_{c,\ell m}, \bs{\omega}(0)) = 0$ for half-integer $\ell$ \cite[ch.~6.7]{barrett2004}.
Similarly, we expect $\omega(\mh{s},\psi) = \omega(-\mh{s},\psi)$ which implies that
$(\bs{\phi}_{c,\ell m}, \bs{\omega}(0)) = 0$ for odd $\ell$. The remaining multiplets correspond to even integer $\ell$, which means we can expect a singlet, a 5-plet, a 9-plet, etc. This explains the multiplet structure of Eq. \eqref{eq:dcsol}.

Several works have considered axially symmetric potentials that can be written
in the form $v(\mh{s}\cdot\mh{s}_c)$, where $\mh{s}_c$ is the axis of symmetry
\cite{stallinga2015, jones2003a}. In this case rotating the potential about the
axis of symmetry commutes with the Smoluchowski operator. Additionally, rotating
the potential by $\pi$ about any axis orthogonal to the symmetry axis also
commutes. We can identify this set of rotations as members of the group
$\mbb{O}(2)$---$2\times 2$ orthogonal matrices. The irreducible representations
of $\mbb{O}(2)$ are one- and two-dimensional \cite{tung1985}, so multiplets can
be at most doublets. Jones demonstrates how perturbing a rotationally symmetric
potential to an axially symmetric potential splits the eigenvalue spectrum into
singlets and doublets \cite{jones2003a}---the original singlet is maintained,
the 5-plet splits into a singlet and two doublets, the 9-plet splits into a
singlet and four doublets, etc.

Note that perturbing the potential also perturbs the eigenfunctions,
so the spherical harmonics will not be eigenfunctions for an axially symmetric
Smoluchowski operator. Perturbing the potential from complete rotational
symmetry will always create eigenfunctions that are not bandlimited, so
bandlimited initial conditions will decay via a superposition of an infinite
number of exponentials.

\subsection{Two-state diffusion and polarized excitation}\label{sec:two-state}
Now we extend our model to a molecule that can be in two states. We
define two functions $w^{\text{(gr)}}(\mh{s},t)$ and $w^{\text{(ex)}}(\mh{s},t)$
as the probabilities that the molecule is in the ground or excited state,
respectively, in orientation $\mh{s}$ at time $t$. We normalize so that
\begin{align}
  \int_{\mbb{S}^2}d\mh{s}\, \left[w^{\text{(gr)}}(\mh{s},t) + w^{\text{(ex)}}(\mh{s},t)\right] = 1.
\end{align}
Next, we define the associated Hilbert-space vectors $\mb{w}^{\text{(gr)}}(t)$
and $\mb{w}^{\text{(ex)}}(t)$, a molecular-species specific
decay rate constant $\kappa^{\text{(d)}}$, and a polarization-dependent
excitation operator $\mc{K}^{\text{(ex)}}_{\mh{p}}$ (parameterized by an
arbitrary polarization state $\mh{p}$) that encodes the orientation-dependent
excitation rate. We assume that the molecule diffuses in the same potential
  while it is in the ground and excited state, so we can model the time-evolution of the molecule as
\begin{align}
  \frac{\partial}{\partial t}
  \begin{bmatrix}
    \mb{w}^{\text{(ex)}}(t)\\
    \mb{w}^{\text{(gr)}}(t)
  \end{bmatrix} =
  \begin{bmatrix}
    \smol - \kappa^{\text{(d)}}& \mc{K}^{\text{(ex)}}_{\mh{p}}\\
    \kappa^{\text{(d)}}&\smol - \mc{K}^{\text{(ex)}}_{\mh{p}}\\
  \end{bmatrix}
  \begin{bmatrix}
    \mb{w}^{\text{(ex)}}(t)\\
    \mb{w}^{\text{(gr)}}(t)
  \end{bmatrix}.
\end{align}
This model assumes that stimulated emission is negligible. This assumption is
justified when the newly excited molecule undergoes fast vibrational relaxation
to a state that is unaffected by stimulated emission from the original
excitation beam. In this two-state model the mean excited-state lifetime is
given by $\tau_e = 1/\kappa^{\text{(d)}}$.

Our goal is to model the observable irradiance
emitted by the molecule as it decays from the excited
state to the ground state. If we expose a detector from $t=t_0$ to $t = t_1$,
then the most we can hope to recover from our measurement is
\begin{align}
  \mb{w} = \int_{t_0}^{t_1}dt\,\kappa^{\text{(d)}}\mb{w}^{\text{(ex)}}(t),\label{eq:intden}
\end{align}
which we call the \textit{point emission density}. Calculating
$\mb{w}$ will almost always require numerics, but we can find
closed-form solutions in several specific cases.

For example, if we assume that diffusion is slow compared to the decay rate constant and
the maximum excitation rate constant
$D\ll \kappa^{\text{(d)}}, \kappa^{\text{(ex)}}_{\text{max}}$, then
we can ignore $\smol$ and write
\begin{align}
  \frac{\partial}{\partial t}
  \begin{bmatrix}
    \mb{w}^{\text{(ex)}}(t)\\
    \mb{w}^{\text{(gr)}}(t)
  \end{bmatrix} =
  \begin{bmatrix}
    -\kappa^{\text{(d)}}& \mc{K}^{\text{(ex)}}_{\mh{p}}\\
    \kappa^{\text{(d)}}& -\mc{K}^{\text{(ex)}}_{\mh{p}}\\
  \end{bmatrix}
  \begin{bmatrix}
    \mb{w}^{\text{(ex)}}(t)\\
    \mb{w}^{\text{(gr)}}(t)
  \end{bmatrix}.
\end{align}
If we excite with coherent light polarized linearly along direction $\mh{p}\in\mbb{S}^2$, then the standard entries of the excitation operator are
\begin{align}
  \left(\mh{e}(\mh{s}),\mc{K}^{\text{(ex)}}_{\mh{p}}\mh{e}(\mh{s}')\right) =  \kappa^{\text{(ex)}}|\mh{p}\cdot\mh{s}|^2\delta(\mh{s} - \mh{s}'), \label{eq:cos}
\end{align}
where $\kappa^{\text{(ex)}}$ is the maximum excitation
rate constant, which is proportional to the intensity of the excitation beam. Rewriting the whole system in a standard basis yields
\begin{align}
  \frac{\partial}{\partial t}
  \begin{bmatrix}
    w^{\text{(ex)}}(\mh{s},t)\\
    w^{\text{(gr)}}(\mh{s},t)
  \end{bmatrix} = 
  \begin{bmatrix}
    -\kappa^{\text{(d)}}& \kappa^{\text{(ex)}}|\mh{p}\cdot\mh{s}|^2\\
    \kappa^{\text{(d)}}& -\kappa^{\text{(ex)}}|\mh{p}\cdot\mh{s}|^2\\
  \end{bmatrix}
  \begin{bmatrix}
    w^{\text{(ex)}}(\mh{s},t)\\
    w^{\text{(gr)}}(\mh{s},t)
  \end{bmatrix}.
\end{align}
If the molecule starts in the ground state $w^{\text{(gr)}}(\mh{s},0) =  1/4\pi$ and $w^{\text{(ex)}}(\mh{s},0) = 0$, then the solution
is given by
\begin{align}
  \begin{bmatrix}
    w^{\text{(ex)}}(\mh{s},t)\\
    w^{\text{(gr)}}(\mh{s},t)
  \end{bmatrix}
   =
  \frac{1}{\kappa^{\text{(ex)}}|\mh{p}\cdot\mh{s}|^2 + \kappa^{\text{(d)}}}
  \begin{bmatrix}
    \kappa^{\text{(ex)}}|\mh{p}\cdot\mh{s}|^2\\
    \kappa^{\text{(d)}}
  \end{bmatrix}
  +
\frac{\kappa^{\text{(ex)}}|\mh{p}\cdot\mh{s}|^2}{\kappa^{\text{(ex)}}|\mh{p}\cdot\mh{s}|^2 + \kappa^{\text{(d)}}}
  \begin{bmatrix}
    -1\\
    1
  \end{bmatrix}
  \exp[-(\kappa^{\text{(ex)}}|\mh{p}\cdot\mh{s}|^2 + \kappa^{\text{(d)}})t].
\end{align}
A particularly interesting result is the steady-state probability of finding the molecule in the excited state
\begin{align}
  w^{\text{(ex)}}(\theta, t \gg \kappa^{\text{(ex)}} + \kappa^{\text{(d)}}) = \frac{\cos^2\theta}{\cos^2\theta + \left[\kappa^{\text{(d)}}/\kappa^{\text{(ex)}}\right]}, \label{eq:angularfree}
\end{align}
where $\theta$ is the angle between $\mh{p}$ and $\mh{s}$.

\begin{figure}[ht]
 \centering
   \centering
   \includegraphics[width = 0.5\textwidth]{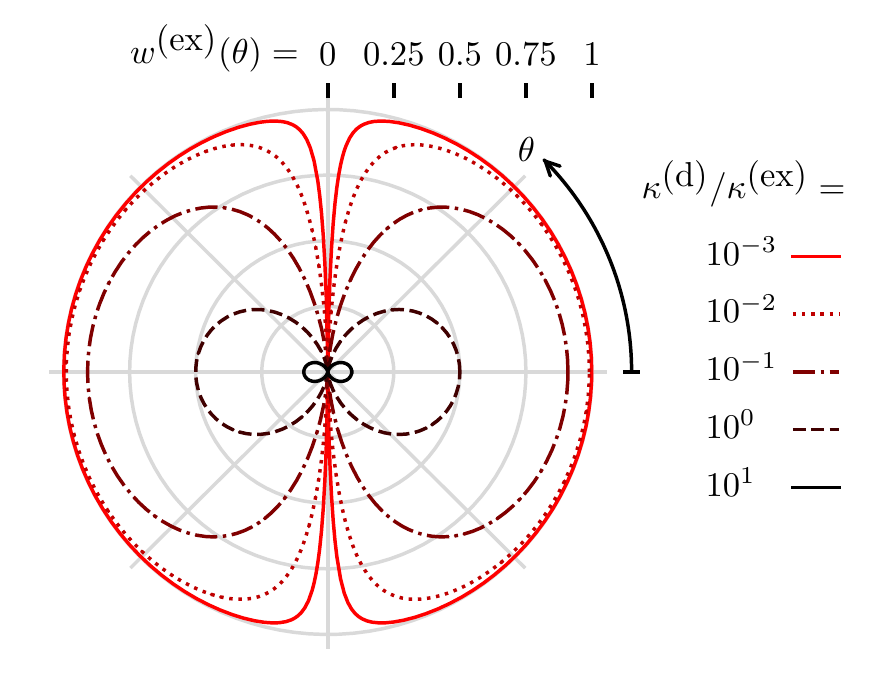}
   \vspace{-2em}
   \caption{Diffusion-free steady-state excited-state probability
     $w^{\text{(ex)}}$ (radius from center) as a function of the angle from the
     incident polarization $\theta$ (clockwise angle from positive $x$ axis) and
     the detection rate to excitation rate ratio
     $\kappa^{\text{(d)}}/\kappa^{\text{(ex)}}$ (color). For weak excitation
     $\kappa^{\text{(d)}}/\kappa^{\text{(ex)}} \gg 1$ the excited-state
     probability is small and only contains low angular-frequency components.
     For strong excitation $\kappa^{\text{(d)}}/\kappa^{\text{(ex)}} \ll 1$ the
     excited state is saturated and contains high angular-frequency components.}
   \label{fig:saturate}
 \end{figure}

Figure \ref{fig:saturate} shows the behavior of Eq. \eqref{eq:angularfree} as
$\theta$ and $\kappa^{\text{(d)}}/\kappa^{\text{(ex)}}$ are varied. For strong
excitation, $\kappa^{\text{(d)}}/\kappa^{\text{(ex)}} \ll 1$ and the excited
state saturates and contains high angular-frequency patterns. These patterns are
directly analogous to the high spatial-frequency patterns generated in
non-linear structured illumination microscopy \cite{gustafsson2005}.

For weak excitation, $\kappa^{\text{(d)}}/\kappa^{\text{(ex)}} \gg 1$ and the $\cos^2\theta$ in the denominator of Eq. \eqref{eq:angularfree} is dwarfed, so the excited state probability is proportional to $\cos^2\theta$. In this limit we are far from saturating the excited state, and the excited-state probability is linear in the excitation power.

\subsection{Two-state diffusion under weak polarized excitation}\label{sec:weak}
In the weak excitation limit, $\kappa^{\text{(d)}}/\kappa^{\text{(ex)}} \gg 1$,
we can approximate the two-state model using an effective one-state model. If
the molecule starts in the ground state and the molecule has diffused to the
steady state then $\mb{w}^{\text{(gr)}}(0) = \bs{\phi}_{\mb{v},0}$. Under weak
excitation the probability of excitation is small, so the ground state
probability will stay approximately constant
$\mb{w}^{\text{(gr)}}(t) \approx \bs{\phi}_{\mb{v},0}$. Our remaining task is to
solve for the excited-state probability, which evolves according to
\begin{align}
  \frac{\partial\mb{w}^{\text{(ex)}}(t)}{\partial t} = -\left(\mc{D}_{\mb{v}} + \kappa^{\text{(d)}}\right)\mb{w}^{\text{(ex)}}(t) + \mc{K}^{\text{(ex)}}_{\mh{p}}\bs{\phi}_{\mb{v},0}.\label{eq:inhomo}
\end{align}
Eq. \eqref{eq:inhomo} is an inhomogeneous system of linear first-order
differential equations. To solve Eq.~\eqref{eq:inhomo}, we start by noticing
that the operator $\left(\smol + \kappa^{\text{(d)}}\right)$ has the same
eigenfunctions as $\smol$ with larger eigenvalues
$\lambda_{\mb{v},i} + \kappa^{\text{(d)}}$. Next, we find the steady-state
solution by setting the left-hand side to zero
\begin{align}
  \mb{w}^{\text{(ex)}}(\infty) = \left(\mc{D}_{\mb{v}} + \kappa^{\text{(d)}}\right)^{-1}\mc{K}^{\text{(ex)}}_{\mh{p}}\bs{\phi}_{\mb{v},0} = \sum_{i=0}^\infty \frac{1}{\lambda_{\mb{v},i} + \kappa^{\text{(d)}}}\left(\frac{\bs{\phi}_{\mb{v},i}}{\bs{\phi}_{\mb{v},0}}, \mc{K}^{\text{(ex)}}_{\mh{p}}\bs{\phi}_{\mb{v},0}\right)\bs{\phi}_{\mb{v},i}.
\end{align}
We can find the homogeneous solution $\mb{w}_h^{\text{(ex)}}(t)$ by ignoring the constant term to find
\begin{align}
  \mb{w}_h^{\text{(ex)}}(t) =  \sum_{i=0}^{\infty}c_{\mb{v},i}\bs{\phi}_{\mb{v},i}\exp\left[-\left(\lambda_{\mb{v},i} + \kappa^{\text{(d)}}\right)t\right].
\end{align}
The complete solution is given by the sum of the homogenous solution and the steady-state solution
\begin{align}
  \mb{w}^{\text{(ex)}}(t) = \mb{w}_h^{\text{(ex)}}(t) + \mb{w}^{\text{(ex)}}(\infty). 
\end{align}
If we begin exposing a detector for a period $t_e$ after the system has reached
a steady state at $t_1 \gg 1/\kappa^{\text{(d)}}$ then the point emission
density is given by
\begin{align}
  \mb{w} &= \int_{t_1}^{t_1 + t_e}dt\, \kappa^{\text{(d)}}\mb{w}^{\text{(ex)}}(\infty)\\
\mb{w} &= \sum_{i=0}^\infty \frac{t_e\kappa^{\text{(d)}}}{\lambda_{\mb{v},i} + \kappa^{\text{(d)}}}\left(\frac{\bs{\phi}_{\mb{v},i}}{\bs{\phi}_{\mb{v},0}}, \mc{K}^{\text{(ex)}}_{\mh{p}}\bs{\phi}_{\mb{v},0}\right)\bs{\phi}_{\mb{v},i}.\label{eq:weaksol}
\end{align}
Eq. \eqref{eq:weaksol} is the main result of this section, and we briefly consider it more closely for cases when diffusion is very slow and very fast.

In the fast diffusion limit \big($\lambda_{\mb{v},i} \gg \kappa^{\text(d)}$ for all
$i>0$\big) all of the terms in Eq. \eqref{eq:weaksol} are negligible except for the $i=0$ term and the result simplifies to
\begin{align}
  \mb{w} \stackrel{\text{(fast)}}{{=}} t_e \left(1, \mc{K}^{\text{(ex)}}_{\mh{p}}\bs{\phi}_{\mb{v},0}\right)\bs{\phi}_{\mb{v},0},\label{eq:weakfast}
\end{align}
which means that the measurable angular distribution is the Boltzmann
distribution weighted by a constant excitation efficiency. Informally, Eq.
\eqref{eq:weakfast} says that a fast diffusing dipole reaches the Boltzmann
distribution before emission, so the emission density is the Boltzmann
distribution multiplied by a constant excitation efficiency.

In the slow diffusion limit \big($\lambda_{\mb{v},i} \ll \kappa^{\text(d)}$ for all
$i$\big), every term in Eq. \eqref{eq:weaksol} contributes and the sum simplifies to 
\begin{align}
  \mb{w} \stackrel{\text{(slow)}}{{=}}
  t_e\mc{K}^{\text{(ex)}}_{\mh{p}}\bs{\phi}_{\mb{v},0}, \label{eq:weakslow}
\end{align}
which means that the measurable angular distribution is the excitation operator
acting on the Boltzmann distribution. Informally, Eq. \eqref{eq:weakslow} says
that a slow diffusing dipole does not rotate before emission, so the point
emission density is the point-wise product of the excitation efficiency function
and the Boltzmann distribution. This situation is the angular analog to linear
structured illumination microscopy \cite{gustafsson2000}, where spatial
diffusion is assumed to be negligible and illumination patterns can be used to
alias high-frequency spatial patterns into the passband of the imaging system.

\subsection{Weak excitation of a free dipole} \label{sec:lastangle}
\begin{figure}[ht]
 \centering
   \centering
   \includegraphics[width = 0.5\textwidth]{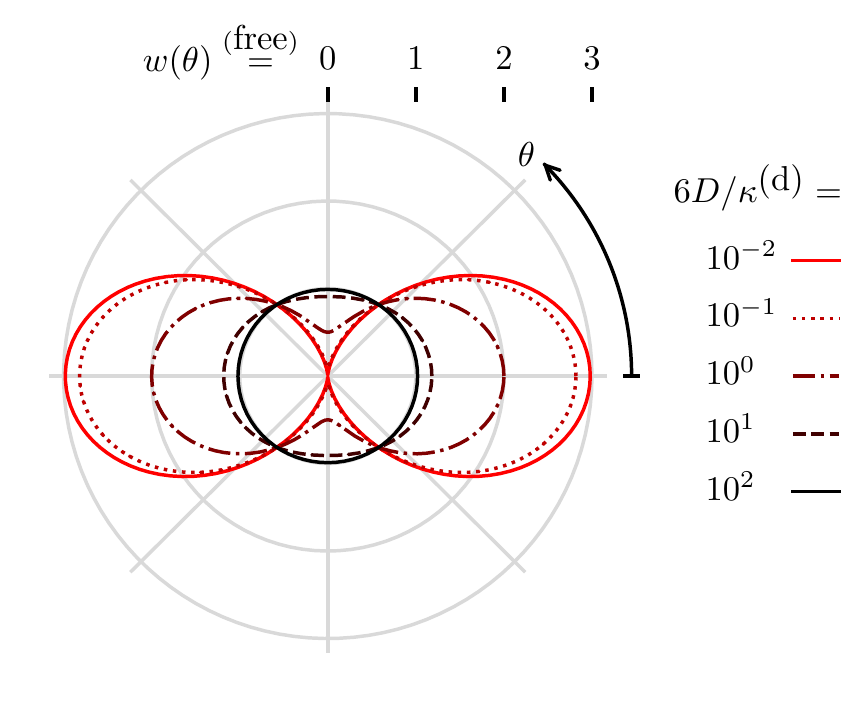}
   \vspace{-2em}
   \caption{Weak-excitation emission density for a free dipole $w$ (radius from
     center) as a function of the angle from the incident polarization $\theta$
     (clockwise angle from positive $x$ axis) and the diffusion rate to decay
     rate constant ratio $6D/\kappa^{\text{(d)}}$ (color). For slow diffusion
     $6D/\kappa^{\text{(d)}} \ll 1$ the emission density is the excitation
     probability, and for fast diffusion $6D/\kappa^{\text{(d)}} \gg 1$ the
     excited emission density is the constant Boltzmann distribution.}
   \label{fig:symrot}
\end{figure}

In the absence of a potential, the eigenvalues become $\lambda_{c,\ell m} = D\ell(\ell+1)$ and the eigenfunctions become $\bs{\phi}_{c,\ell m} = \mh{E}_{\ell m}/\sqrt{4\pi}$ (see Section \ref{sec:symmetric}). Plugging these into Eq. \eqref{eq:weaksol} yields
\begin{align}
\mb{w} &\stackrel{\text{(free)}}{{=}} \sum_{\ell=0}^\infty\sum_{m=-\ell}^\ell \frac{t_e\kappa^{\text{(d)}}}{D\ell(\ell+1) + \kappa^{\text{(d)}}}\left(\mh{E}_{\ell m}, \mc{K}^{\text{(ex)}}_{\mh{p}}/4\pi\right)\mh{E}_{\ell m}. \label{eq:spheresym}
\end{align}
For linearly polarized coherent illumination we can write Eq. \eqref{eq:spheresym} in the standard basis as
\begin{align}
w(\theta) &\stackrel{\text{(free)}}{{=}} \frac{t_e\kappa^{\text{(ex)}}}{12\pi}\left[1 +\frac{3\cos^2\theta -1}{1 + [6D/\kappa^{\text{(d)}}]}\right], \label{eq:spheresym3}
\end{align}
where $\theta$ is the angle between $\mh{p}$ and $\mh{s}$. Figure
\ref{fig:symrot} shows the behavior of Eq. \eqref{eq:spheresym3} as $\theta$ and
$6D/\kappa^{\text{(d)}}$ are varied. For slow diffusion
($6D/\kappa^{\text{(d)}} \ll 1$) the point emission density is identical
to the excitation probability $|\mh{p}\cdot\mh{s}|^2$, and for fast diffusion
($6D/\kappa^{\text{(d)}} \gg 1$) the point emission density is the
constant Boltzmann distribution.

Notice that the infinite sum in Eq. \eqref{eq:spheresym} reduces to two non-zero
terms in Eq. \eqref{eq:spheresym3} because coherent polarized illumination
excites the (constant) Boltzmann distribution into a linear combination of six
eigenfunctions that have only two distinct eigenvalues. For asymmetric
potentials, coherent polarized illumination will excite the Boltzmann
distribution into a linear combination of an infinite number of eigenfunctions,
so the solution will contain an infinite number of terms.

\subsection{Spatio-angular emission densities}
So far we have been considering point emission densities
$\mb{w} \in \mbb{L}_2(\mbb{S}^2)$ for single molecules at a single point in
space. In this section we will extend our discussion to ensembles of molecules
and three-dimensional angular emission densities represented by vectors
$\mb{f} \in \mbb{L}_2(\mbb{R}^3\times\mbb{S}^2)$.

We start by defining a spatio-angular dynamics model similar to the angular
dynamics model in Section \ref{sec:two-state}. First, we define a pair of
functions $f^{\text{(gr)}}(\ro,\so,t)$ and $f^{\text{(ex)}}(\ro,\so,t)$ as the
number of molecules in the ground and excited states, respectively, at position
$\ro\in\mbb{R}^3$, in orientation $\mh{s}\in\mbb{S}^2$, and at time $t$ (per unit
volume, solid angle, and time). These unnormalized functions are related to the
normalized functions we considered earlier by
\begin{align}
  f^{\text{(gr)}}(\ro,\so,t) &= \rho(\ro,t)w^{\text{(gr)}}(\ro,\so,t),\\
  f^{\text{(ex)}}(\ro,\so,t) &= \rho(\ro,t)w^{\text{(ex)}}(\ro,\so,t),
\end{align}
where $\rho(\ro,t)$ is an orientation-independent \textit{spatial density}---the
number of fluorescent molecules per unit volume at point $\ro$. We also define
the associated Hilbert-space vectors $\mb{f}^{\text{(gr)}}(t)$ and
$\mb{f}^{\text{(ex)}}(t)$. Next, we define a spatio-angular potential $\bs{v}$,
a Smoluchowski operator $\mathscr{D}_{\bs{v}}$ that models how molecules diffuse
in the spatio-angular potential $\bs{v}$, a spatio-angular excitation operator
$\mathscr{K}^{\text{(ex)}}_{\mh{p}}$, and a decay rate constant
$\kappa^{\text{(d)}}$. With these definitions we can model the spatio-angular
populations with
\begin{align}
  \frac{\partial}{\partial t}
  \begin{bmatrix}
    \mb{f}^{\text{(ex)}}(t)\\
    \mb{f}^{\text{(gr)}}(t)
  \end{bmatrix} =
  \begin{bmatrix}
    \mathscr{D}_{\bs{v}} - \kappa^{\text{(d)}}& \mathscr{K}^{\text{(ex)}}_{\mh{p}}\\
    \kappa^{\text{(d)}}&\mathscr{D}_{\bs{v}} - \mathscr{K}^{\text{(ex)}}_{\mh{p}}\\
  \end{bmatrix}
  \begin{bmatrix}
    \mb{f}^{\text{(ex)}}(t)\\
    \mb{f}^{\text{(gr)}}(t)
  \end{bmatrix}.\label{eq:samodel}
\end{align}
We are interested in the spatio-angular emission density during an exposure from $t_0$ to $t_1$ given by
\begin{align}
  \mb{f} = \int_{t_0}^{t_1}dt\,\kappa^{\text{(d)}}\mb{f}^{\text{(ex)}}(t).\label{eq:intden2}
\end{align}

If spatial diffusion is negligible, then the spatio-angular model in Eq.
\eqref{eq:samodel} decouples into an angular model at each point weighted by a
time-independent spatial density $\rho(\ro)$, so we can write
\begin{align}
  f(\ro,\so) = \rho(\ro)w(\ro,\so).\label{eq:slowspatial}
\end{align}
For each spatial point $\ro$ we can use the angular solutions developed in
Sections \ref{sec:two-state}--\ref{sec:lastangle} to calculate $w(\ro,\so)$.

\subsection{Spatio-angular imaging operator}
In this section we complete our imaging model by finding the mapping between the
spatio-angular emission density $\mb{f}$ and the data we measure $\mb{g}$. It will be convenient to choose a basis for $\mb{f}$ that splits the object-space spatial coordinates into a one-dimensional longitudinal coordinate $\ropar$ aligned with the optical axis of the microscope, and a two-dimensional transverse coordinate $\roperp$. More specifically, we express $\mb{f}$ in the following basis
\begin{align}
  f(\roperp, \ropar, \so) = (\mh{e}(\roperp)\mh{e}(\ropar)\mh{e}(\mh{s}), \mb{f}).
\end{align}
Next, we model the irradiance measured at each point on a planar detector with the
function $g(\rd)$ with $\rd \in \mbb{R}^2$ or its associated Hilbert-space
vector $\mb{g} \in \mbb{L}_2(\mbb{R}^2)$. Finally, we model the mapping between emission
densities and data with a Hilbert-space operator $\mc{H}:\mbb{L}_2(\mbb{R}^3\times\mbb{S}^2)\rightarrow \mbb{L}_2(\mbb{R}^2)$ that acts on $\mb{f}$
\begin{align}
  \mb{g} = \mc{H}\mb{f}.
\end{align}

Several works \cite{nov2006, lew2013, backer2014, stallinga2015, chandler2019b} have calculated the standard entries of $\mc{H}$ for an aplanatic $4f$ optical system with a paraxial tube lens and unit magnification (or demagnified coordinates) as 
\begin{align}
  h(\rd,\roperp,\ropar,\so) \equiv \left(\mh{e}(\rd), \mc{H}\mh{e}(\roperp)\mh{e}(\ropar)\mh{e}(\so)\right) = \sum_{i=0,1}|c_i(\rd-\roperp,\ropar,\so)|^2,
\end{align}
where 
\begin{align}
  c_i(\mb{r}^\perp,\ropar,\so) &= \int_{\mbb{R}^2}d\bs{\tau}\,C_i(\bs{\tau},\ropar,\so)\exp[i2\pi\bs{\tau}\cdot\mb{r}^\perp]
\end{align}
is the $i$th component of the dipole coherent spread function,
\begin{align}
    C_i(\bs{\tau},\ropar,\so) &=A(\bs{\tau})\Phi(\bs{\tau},\ropar)
\sum_{j=0,1,2}g_{ij}(\bs{\tau})s_j\end{align}
is the $i$th component of the dipole coherent transfer function,
\begin{align}
  A(\bs{\tau}) = (1 - |\bs{\tau}|^2)^{-1/4}\Pi\left(|\bs{\tau}|/\nu_c\right)\label{eq:highNA0}
\end{align}
is the aplanatic apodization function with full width $\nu_c = 2\text{NA}/\lambda$,
\begin{align}
  \Phi(\bs{\tau},\ropar) = \exp\left[i2\pi\ropar\sqrt{\nu_m^2 - |\bs{\tau}|^2}\right]
\end{align}
is a phase-encoding function with $\nu_m = n_0/\lambda$, the functions $g_{ij}(\bs{\tau})$ model the $i$th field components in the pupil plane created by the $j$th component of a dipole
\begin{alignat}{2}
  g_{00}(\bs{\tau}) &= \sin^2\phi_{\tau}  + \cos^2\phi_{\tau}\sqrt{1-|\bs{\tau}|^2},\qquad &&g_{10}(\bs{\tau}) = \frac{1}{2}\sin(2\phi_{\tau})\left(\sqrt{1 - |\bs{\tau}|^2} -1\right),\nonumber\\
  g_{01}(\bs{\tau}) &= \frac{1}{2}\sin(2\phi_{\tau})\left(\sqrt{1 - |\bs{\tau}|^2} -1\right), \qquad &&g_{11}(\bs{\tau}) = \cos^2\phi_\tau  + \sin^2\phi_\tau\sqrt{1 - |\bs{\tau}|^2},\nonumber\\
  g_{02}(\bs{\tau}) &= |\bs{\tau}|\cos\phi_\tau,
  \qquad &&g_{12}(\bs{\tau}) = |\bs{\tau}|\sin\phi_\tau,\label{eq:highNA1}
\end{alignat}
and $s_j$ is the $j$th component of the dipole unit vector $\so$. This model is
accurate for objectives with arbitrarily high numerical apertures (provided the
objective is free from aberration and satisfies the aplanatic condition). Eqs.
\eqref{eq:highNA0}--\eqref{eq:highNA1} model the apodization, phase shifts, and
directional electric fields in high-NA optical systems, and paraxial models have
been constructed by approximating these functions with low-order polynomials
\cite{chandler2019b}. Modeling a mask in the back aperture of the objective can
be accomplished by modifying the amplitude $A$ or phase $\Phi$ functions.
Following Stallinga \cite{stallinga2015}, we can rewrite the standard entries in
a form that is more efficient for computation
\begin{align}
  h(\rd,\roperp,\ropar,\so) = \sum_{j,j'=0,1,2}B_{jj'}(\rd-\roperp,\ropar)s_js_{j'},
\end{align}
where
\begin{align}
  B_{jj'}(\mb{r}^\perp,\ropar) = \sum_{i=0,1} \beta_{ij}(\mb{r}^\perp,\ropar)\beta_{ij'}^*(\mb{r}^\perp,\ropar),
\end{align}
and
\begin{align}
  \beta_{ij}(\mb{r}^\perp,\ropar) = \int_{\mbb{R}^2}d\bs{\tau}\,A(\bs{\tau})\Phi(\bs{\tau},\ropar)g_{ij}(\bs{\tau})\exp[i2\pi\bs{\tau}\cdot\mb{r}^\perp].
\end{align}
For general amplitude and phase masks, six Fourier transforms need to be computed for each defocus position. If the amplitude and phase masks are radial ($A(\bs{\tau})$ and $\Phi(\bs{\tau},\ropar)$ are independent of $\phi_{\tau}$), then we can exploit the following symmetries
\begin{align}
  \beta_{00}(\mb{r}^\perp,\ropar) &= \beta_{11}(\mc{R}_{\pi/2}\mb{r}^\perp,\ropar), \\
  \beta_{01}(\mb{r}^\perp,\ropar) &= \beta_{10}(\mb{r}^\perp,\ropar), \\
  \beta_{02}(\mb{r}^\perp,\ropar) &= \beta_{12}(\mc{R}_{\pi/2}\mb{r}^\perp,\ropar),
\end{align}
where $\mc{R}_{\pi/2}$ is an operator that rotates transverse coordinates by $\pi/2$, and only compute three Fourier transforms per defocus position.

We can calculate the entries of $\mc{H}$ in other bases by relating them to the
standard entries. Choosing the spherical harmonics for the input basis is
convenient because it allows us to exploit the angular bandlimit of the imaging
system and work in an orthonormal basis. Calculating the entries in this basis
yields
\begin{align}
  H_{\ell m}(\rd,\roperp,\ropar) &\equiv \left(\mh{e}(\rd), \mc{H}\mh{e}(\roperp)\mh{e}(\ropar)\mh{E}_{\ell m}\right)\nonumber\\
                                 &= \int_{\mbb{S}^2}d\so h(\rd,\roperp,\ropar,\so)Y_{\ell m}(\so)\nonumber\\
                                 &= \sum_{j,j'=0,1,2}\left[\int_{\mbb{S}^2}d\so\, Y_{\ell m}(\so)s_js_{j'}\right] B_{jj'}(\rd-\roperp,\ropar)\nonumber\\
                                 &= \frac{4\pi}{3}\sum_{j,j'=0,1,2}G_{\ell 11}^{m\epsilon_j\epsilon_{j'}}B_{jj'}(\rd-\roperp,\ropar),\label{eq:matrixel}
\end{align}
where $\epsilon_0 = 1$, $\epsilon_1 = -1$, $\epsilon_2 = 0$, and
\begin{align}
  G_{\ell\ell'\ell''}^{mm'm''} = \int_{\mbb{S}^2}d\mh{s}\,Y_{\ell m}(\mh{s})Y_{\ell' m'}(\mh{s})Y_{\ell'' m''}(\mh{s})
\end{align}
are the real Gaunt coefficients \cite{homeier1996, shirdhonkar2005}. The Gaunt coefficients $G_{\ell 11}^{mm'm''}$ are only nonzero for $\ell = 0,2$, which means that
$\mc{H}$ only transmits six angular components. 

\section{Results}\label{sec:results}
To demonstrate our model we will specify a geometric phantom under three
different limits (fast angular diffusion, slow angular diffusion, and free
dipoles), specify an imaging system, then simulate the irradiance patterns
generated by the phantom under these three limits.

\subsection{Phantom}
We begin by choosing the following family of angular potentials
\begin{align}
  v_0(\so;\theta) = -V_0(\so\cdot[\mh{y}\sin\theta + \mh{z}\cos\theta])^2,
\end{align}
where $\so$ is the object-space angular variable, $\theta$ is the angle between
the symmetry axis and the $\mh{z}$ axis in the $\mh{y}$-$\mh{z}$ plane, and
$V_0$ is a positive constant. The corresponding Boltzmann distributions are
\begin{align}
  \phi_0(\so;\theta) = Z^{-1}\exp[V_0\beta (\so\cdot[\mh{y}\sin\theta + \mh{z}\cos\theta])^2], 
\end{align}
which are Watson distributions with mean orientations
$[\mh{y}\sin\theta + \mh{z}\cos\theta]$ and concentration parameter $V_0\beta$
\cite{mardia2000}. For our simulations we fix the concentration parameter to
$V_0\beta = 4$. This family of dipole distributions has its mean orientation in
the $\mh{y}$-$\mh{z}$ plane, but the dipoles are not restricted to this plane.
Next, we define a spatio-angular potential as
\begin{align}
  v(\ro,\so) = v_0(\so;[\pi/4]\ro\cdot\mh{y}),
\end{align}
which consists of distributions with mean orientations that change with $\ro\cdot\mh{y}$. We assume that spatial diffusion is negligible, so we can write
the spatio-angular equilibrium distribution as
\begin{align}
  \phi(\ro,\so) = \phi_0(\so;[\pi/4]\ro\cdot\mh{y}),
\end{align}
and we can choose a time-independent spatial density
\begin{align}
  \rho(\ro) = \sum_{i=0}^{2} \sum_{j=0}^{2} \delta(\ro - i\mh{x} - j\mh{y} - [j/4]\mh{z}).\label{eq:density}
\end{align}
The geometric phantom consists of nine labeled points in a three-dimensional
grid measured in $\mu$m. The three rows of points are increasingly defocused (0,
0.25, and 0.5 $\mu$m of defocus), and the three columns of points have mean
orientations that are increasingly tilted away from the $\mh{z}$ axis towards
the $\mh{y}$ axis (0, $\pi/4$, and $\pi/2$ radians between the mean orientation
and the $\mh{z}$ axis). Finally, we illuminate the sample with coherent light
linearly polarized along the $\mh{p}$ axis with standard entries
\begin{align}
\left(\mh{e}(\mh{s}),\mc{K}^{\text{(ex)}}_{\mh{p}}\mh{e}(\mh{s}')\right) =  \left(\mh{p}\cdot\mh{s}'\right)^2\delta(\mh{s} - \mh{s}').\label{eq:exc2}
\end{align}

Now that we have specified the geometry of our phantom, we will calculate the
emission densities under three limits (fast angular diffusion, slow angular
diffusion, and free dipoles). Plugging Eqs. \eqref{eq:density} and
\eqref{eq:exc2} into Eqs. \eqref{eq:weakfast} and \eqref{eq:slowspatial} yields
the following emission density for weak excitation of dipoles undergoing fast
angular diffusion
\begin{align}
  f^{\mh{p}}_{\text{(fast)}}(\ro,\so) = \rho(\ro)\left[\int_{\mbb{S}^2 }d\mh{s}\,(\mh{p}\cdot\mh{s})^2\phi(\ro,\mh{s})\right]\phi(\ro,\so).\label{eq:phantomfast}
\end{align}
Using Eq. \eqref{eq:weakslow} instead of Eq. \eqref{eq:weakfast} yields the following emission density for weak excitation of dipoles undergoing slow angular diffusion
\begin{align}
  f^{\mh{p}}_{\text{(slow)}}(\ro,\so) = \rho(\ro)\left[(\mh{p}\cdot\so)^2\phi(\ro,\so)\right].\label{eq:phantomslow}
\end{align}

For our final phantom we consider free dipoles (no angular potential) with a
spatially varying ratio $6D/\kappa^{\text{(d)}}$. We modify Eq. \eqref{eq:spheresym3} to create the emission density
\begin{align}
  f^{\mh{p}}_{\text{(free)}}(\ro,\so) = \rho(\ro)\left[1 +\frac{3(\so\cdot\mh{p})^2 -1}{1 + 10^{(\ro\cdot\mh{y}) - 1}}\right],\label{eq:phantomfree}
\end{align}
where the factor $10^{(\ro\cdot\mh{y}) - 1}$ models a position-dependent
rotational mobility in the phantom.

\subsection{Imaging system}
To simulate our imaging system, we start with a phantom $f(\ro,\so)$, change to a basis of spherical harmonics using 
\begin{align}
  F_{\ell m}(\ro) = \int_{\mbb{S}^2}d\so f(\ro,\so)Y_{\ell m}(\so),
\end{align}
then simulate the data using 
\begin{align}
  g(\rd) = \lmsum\int_{\mbb{R}^2}d\ro H_{\ell m}(\rd,\ro)F_{\ell m}(\ro),
\end{align}
where the matrix elements $H_{\ell m}(\rd,\ro)$ can be calculated with Eq. \eqref{eq:matrixel}. Note that $H_{\ell m}(\rd,\ro) = 0$ for $\ell > 2$, so we only
need to calculate $F_{\ell m}(\ro)$ for $\ell \leq 2$---six total entries. 

We choose NA = 1.4, $\lambda = 500$ nm, and $n_0$ = 1.5. We sample and plot the scaled irradiance at 20$\times$ the Nyquist rate, $\Delta x$ = 1/[20(2$\nu_c$)], so the irradiance patterns are free of aliasing.

\subsection{Simulated irradiance patterns}
\begin{figure}[ht]
 \centering
   \includegraphics[width = 1.0\textwidth]{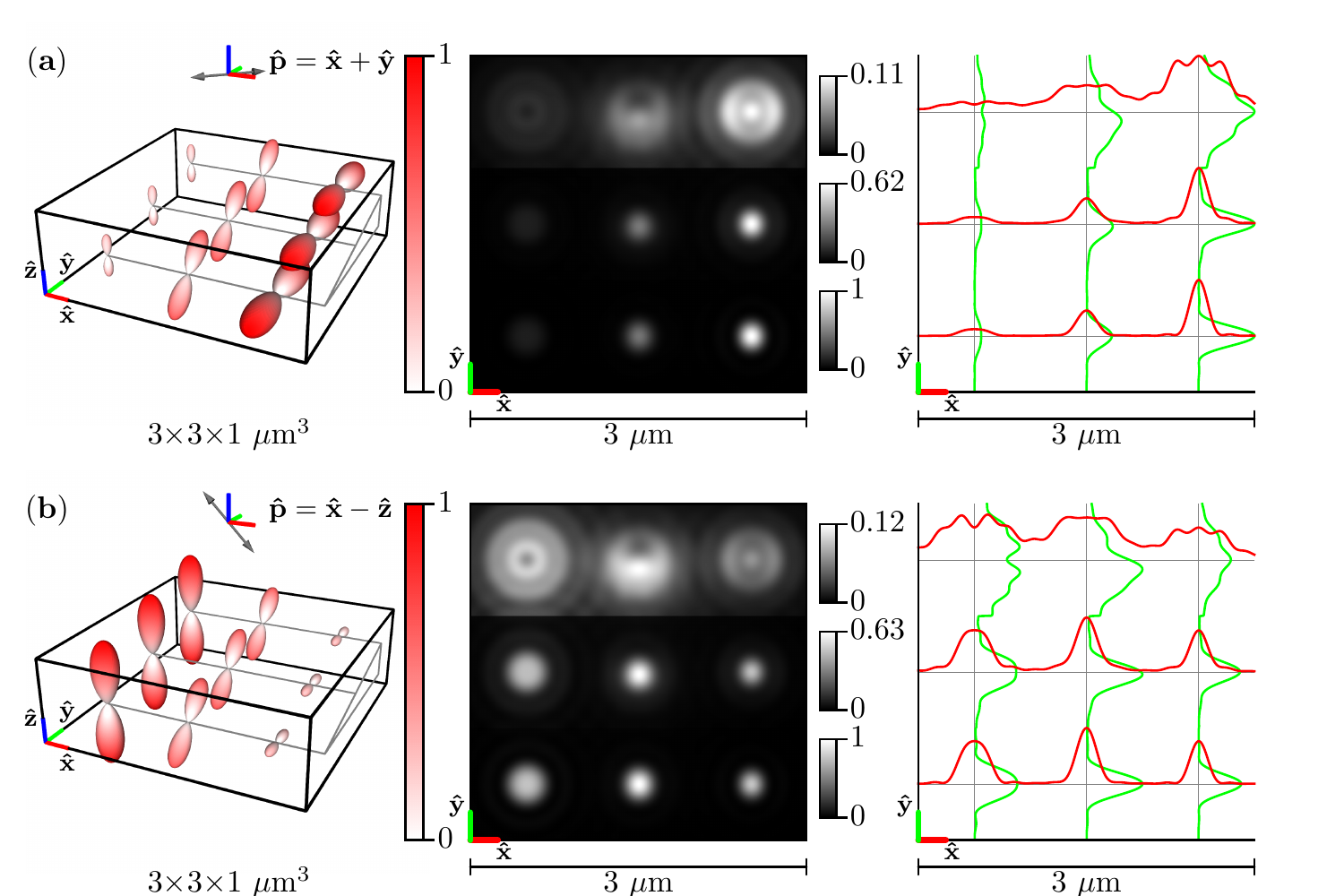}
   \caption{\textbf{Left:} A spatio-angular phantom undergoing fast angular diffusion---see Eq. \eqref{eq:phantomfast}---under illumination by \textbf{(a)} $[\mh{x}+\mh{y}]$ polarized light and \textbf{(b)} $[\mh{x}-\mh{z}]$ polarized light. The phantom consists of nine point sources with varying defocus (rows) and mean orientation (columns). The radius and color of each glyph encode the value of the emission density $f^{\mh{p}}_{\text{(fast)}}(\ro,\so)$. \textbf{Center:} Irradiance patterns for an imaging system with NA = 1.4, $\lambda$ = 500 nm, and $n_0 = 1.5$ sampled at 20$\times$ the Nyquist rate. Each row is individually normalized as indicated by the color bars. \textbf{Right:} Horizontal (red) and vertical (green) profiles through the irradiance pattern.}
   \label{fig:output0}
\end{figure}

Figure \ref{fig:output0} shows $f^{\mh{p}}_{\text{(fast)}}(\ro,\so)$ under two
illumination polarizations ($\mh{p} = \mh{x} + \mh{y}$ and
$\mh{p} = \mh{x} - \mh{z}$), their images, and profiles through each image. Fast-diffusing dipoles reach their Boltzmann distribution before decaying, so the
emission densities in Fig. \ref{fig:output0} are Watson-distributed and
rotationally symmetric about a mean axis. The emission densities are weighted by
constant excitation efficiencies (see Eqs. \eqref{eq:weakfast} and
\eqref{eq:phantomfast}), so each Watson distribution is scaled by a constant
factor. Distributions that have more dipoles aligned parallel to the
polarization direction are excited most efficiently, and the emission densities
in Fig. \ref{fig:output0} are scaled to represent this fact.

As expected, in-focus distributions (the bottom row of distributions in Figs.
\ref{fig:output0}(\textbf{a}) and \ref{fig:output0}(\textbf{b})) generate the
brightest and most rotationally symmetric irradiance patterns, while defocused
distributions spread the irradiance over a larger area on the detector and
oblique defocused distributions display asymmetric irradiance patterns (the top
row and center column is asymmetric in the $\mh{y}$ direction). Notably,
fast-diffusing dipoles under different polarized illuminations create irradiance
patterns with different scales and the same shape.

\begin{figure}[ht]
 \centering
   \includegraphics[width = 1.0\textwidth]{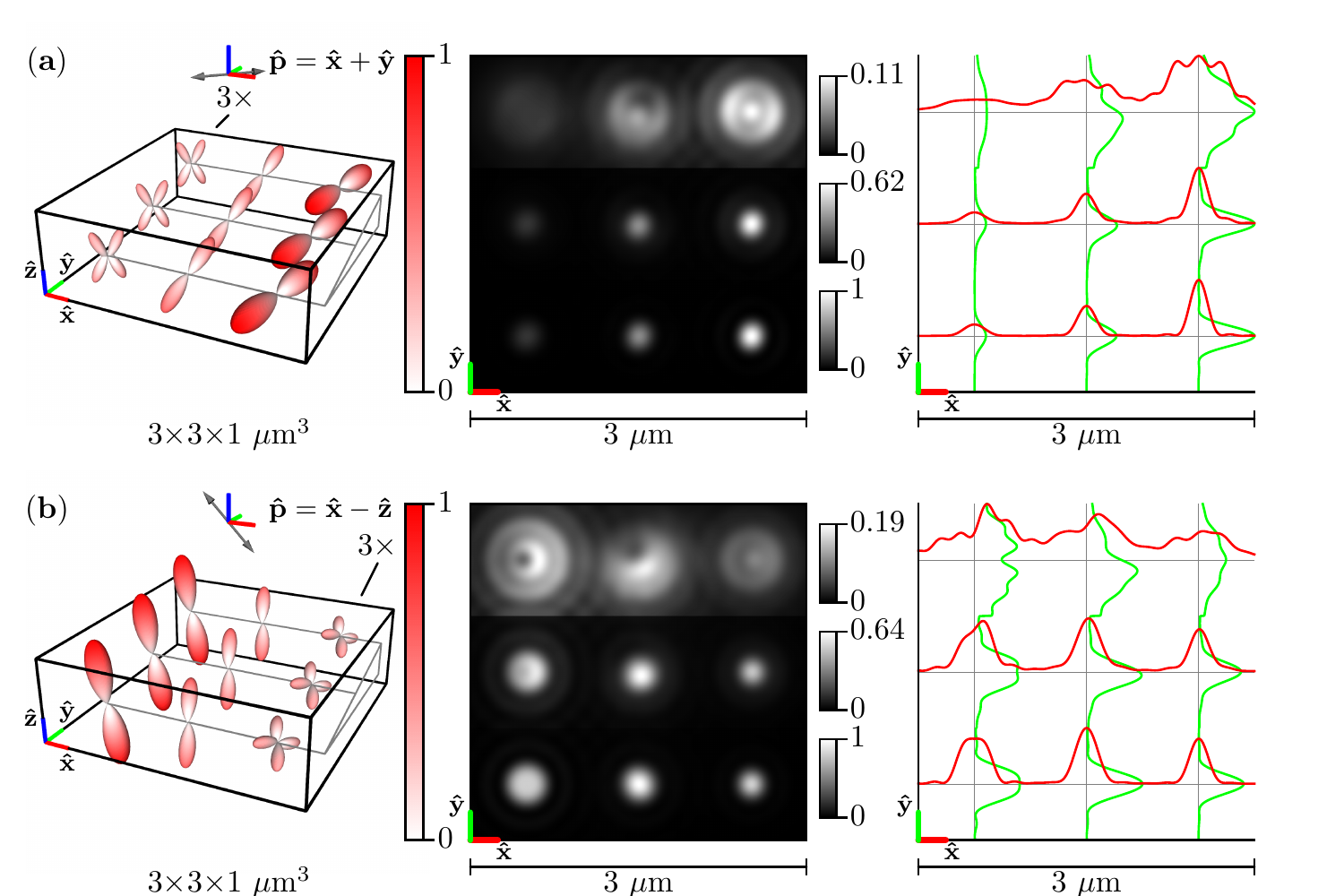}
   \caption{\textbf{Left:} A spatio-angular phantom undergoing slow angular
     diffusion---see Eq. \eqref{eq:phantomslow}---under illumination by
     \textbf{(a)} $[\mh{x}+\mh{y}]$ polarized light and \textbf{(b)}
     $[\mh{x}-\mh{z}]$ polarized light. The phantom consists of nine point
     sources with varying defocus (rows) and mean orientation (columns). The
     radius and color of each glyph encode the value of the emission density
     $f^{\mh{p}}_{\text{(slow)}}(\ro,\so)$, and the left column (\textbf{a}) or
     right column (\textbf{b}) of glyphs are magnified by $3\times$ for
     visualization purposes. \textbf{Center:} Irradiance patterns for an
     imaging system with NA = 1.4, $\lambda$ = 500 nm, and $n_0 = 1.5$ sampled
     at 20$\times$ the Nyquist rate. Each row is individually normalized as
     indicated by the color bars. \textbf{Right:} Horizontal (red) and vertical
     (green) profiles through the irradiance pattern.}
   \label{fig:output1}
\end{figure}

Figure \ref{fig:output1} shows the same results as Fig. \ref{fig:output0} but in
the slow diffusion limit $f^{\mh{p}}_{\text{(slow)}}(\ro,\so)$. Slow-diffusing
dipoles do not rotate before emission, so the emission density is the point-wise
product of the excitation efficiency function and the Boltzmann distribution
(see Eqs. \eqref{eq:weakslow} and \eqref{eq:phantomslow}). Importantly, this
means that the emission densities are not rotationally symmetric (the point-wise
product of two rotationally symmetric functions is not always rotationally
symmetric). This asymmetry is especially apparent for Watson distributions with
mean directions that are perpendicular to the polarization direction (the left
column in Fig. \ref{fig:output1}(\textbf{a}) and the right column in Fig.
\ref{fig:output1}(\textbf{b})). In addition to rotational asymmetry, slow-diffusing dipole emission density maxima are tilted towards the excitation
polarized direction (see the right column in Fig. \ref{fig:output1}(\textbf{a})
and the left column in Fig. \ref{fig:output1}(\textbf{b})), which is due to the
point-wise product of the excitation efficiency function and the Boltzmann
distribution.

The slow-diffusing dipoles in Fig. \ref{fig:output1} display more asymmetric
irradiance patterns than the fast-diffusing dipoles in Fig. \ref{fig:output0}.
Perhaps surprisingly, defocused slow-diffusing dipoles display irradiance
asymmetry along both the $\mh{x}$ and $\mh{y}$ directions (see the top row and
center column of Figs. \ref{fig:output1}(\textbf{a}) and
\ref{fig:output1}(\textbf{b})) despite the fact that the Watson distribution
means are in the $\mh{y}$--$\mh{z}$ plane. This effect is a direct consequence
of the excitation polarization---the emission density maxima are tilted towards
the polarization axis, which gives the emission density maxima $\mh{x}$,
$\mh{y}$, and $\mh{z}$ components.
 
\begin{figure}[ht]
 \centering
   \includegraphics[width = 1.0\textwidth]{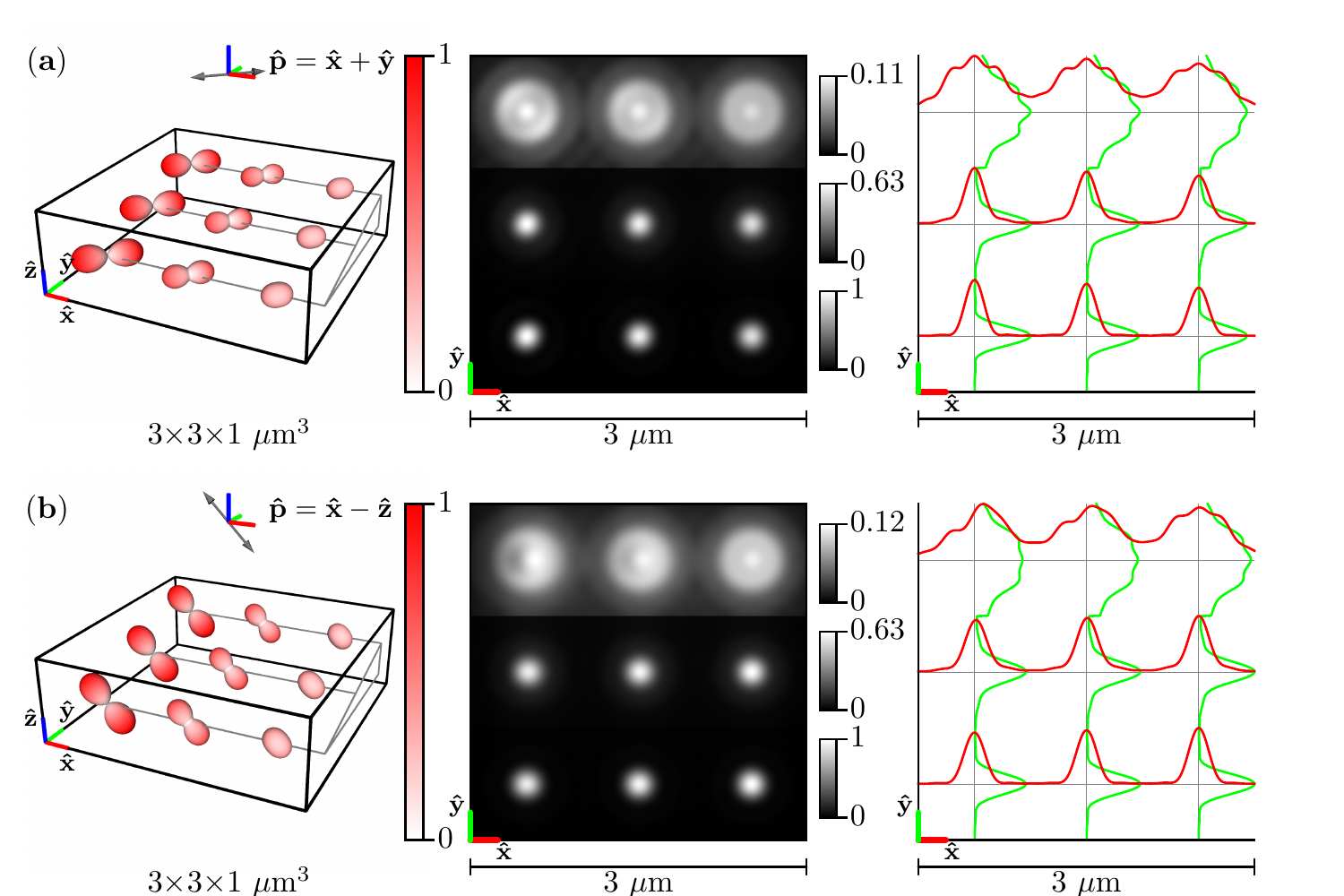}
   \caption{\textbf{Left:} A spatio-angular phantom consisting of free dipoles
     ---see Eq. \eqref{eq:phantomfree}---under illumination by \textbf{(a)}
     $[\mh{x}+\mh{y}]$ polarized light and \textbf{(b)} $[\mh{x}-\mh{z}]$
     polarized light. The phantom consists of nine point sources with varying
     defocus (rows) and ratio of the diffusion coefficient to the decay rate
     $D/\kappa^{\text{(d)}}$ (columns). The radius and color of each glyph
     encode the value of the emission density
     $f^{\mh{p}}_{\text{(free)}}(\ro,\so)$. \textbf{Center:} Irradiance patterns
     for an imaging system with NA = 1.4, $\lambda$ = 500 nm, and $n_0 = 1.5$
     sampled at 20$\times$ the Nyquist rate. Each row is individually normalized
     as indicated by the color bars. \textbf{Right:} Horizontal (red) and
     vertical (green) profiles through the irradiance pattern.}
   \label{fig:output2}
 \end{figure}

Figure \ref{fig:output2} shows the results for the phantom of freely diffusing
dipoles $f^{\mh{p}}_{\text{(free)}}(\ro,\so)$. For slow-diffusing free
dipoles (left column), the emission density is identical to the excitation
efficiency function, while for fast-diffusing free dipoles (right column), the
emission density is nearly uniform. The irradiance patterns are similar for
slow- and fast-diffusing free dipoles under different polarizations, but [$\mh{x}-\mh{z}$]-polarized illumination of slow-diffusing defocused dipoles create asymmetric irradiance patterns (top row, left column of Fig. \ref{fig:output2}(\textbf{b})).
 
\section{Discussion and Conclusions}\label{sec:discussion}
\subsection{When are diffusion decays multi-exponential, and when does it matter?}
Existing works use monoexponential decays to model angular diffusion
\cite{lew2013, backer2015, stallinga2015}---they assume that angular diffusion
within a potential can be characterized by a single rotational relaxation time.
We have shown that this assumption is only justified when the initial condition
is a linear combination of eigenfunctions of the Smoluchowski operator that
share a single non-zero eigenvalue. The monoexponential assumption is true for
weak linearly polarized excitation of molecules in solution---the initial
conditions and the eigenfunctions of the Smoluchowski operator are linear
combinations of the $\ell = 0$ and $\ell=2$ spherical harmonics that share a
single non-zero eigenvalue. Perturbing the angular potential will change the
eigenfunctions of the Smoluchowski operator and lead to multi-exponential
decays.

In the limits of weak excitation and fast diffusion
($\lambda_{\mb{v},i} \gg \kappa^{\text{(d)}}$ for all $i > 0$) or slow diffusion
($\lambda_{\mb{v},i} \ll \kappa^{\text{(d)}}$), our results agree with the
literature that makes the monoexponential assumption \cite{lew2013, backer2015,
  stallinga2015}. Weakly excited fast-diffusing dipoles reach their Boltzmann
distribution before emission, so the emission density is the Boltzmann
distribution multiplied by a constant excitation efficiency factor. Meanwhile,
weakly excited slow-diffusing dipoles do not rotate before emission, so the
emission density is the pointwise product of the excitation efficiency function
and the Boltzmann distribution.

Our predictions diverge from the monoexponential literature \cite{lew2013,
  backer2015, stallinga2015} in the intermediate regime
($\lambda_{\mb{v},i} \approx \kappa^{\text{(d)}}$ for any $i > 0$), and the
differences can be dramatic. To choose an adversarial example, consider an
angular potential with two wells separated by a large but finite potential
barrier. If molecules within one well are excited then they can diffuse quickly
within that well, but they will take a long time to diffuse to the other well.
Clearly, multiple diffusion times are needed to characterize the imaging
process.

Multiple diffusion times are needed to characterize simpler angular potentials,
too. Consider the widely used ``wobble-in-a-cone'' model with a single molecule
that is initially oriented at the edge of the cone. In Sections
\ref{sec:arbitrary}--\ref{sec:symmetric} we used differential equations and
group theory to argue that the angular decay will be multi-exponential, but we
can understand the argument qualitatively by approximating the diffusion as a
discrete random walk of a single molecule. Initially, the dipole can move in
three directions each with probability $\approx$1/3---clockwise,
counterclockwise, and towards the center of the cone. Later, the molecule will
be away from the edge of the cone, and the molecule can move in four directions
with probability $\approx$1/4. Therefore, molecules at the edge will move away
from their initial condition faster on average than molecules away from the
edge, so a single diffusion time is insufficient to characterize diffusion with
a cone.

\subsection{Is angular structured illumination different from polarized illumination?}
We have used the term ``angular structured illumination'' instead of ``polarized
illumination'' throughout this work for two reasons. First, unpolarized light
can have angular structure---an unpolarized plane wave does not excite dipoles
parallel to its propagation direction. Although we have focused on using
polarized light to alias high angular frequency information into the passband,
unpolarized light can be used to the same effect (albeit with less efficient
aliasing than polarized light). Second, ``angular structured illumination''
highlights the deep similarity with spatial structured illumination. Readers
familiar with spatial structured illumination can apply their intuition to
angular structured illumination techniques, and many existing spatial techniques
have direct angular analogs.

\subsection{How many angular components can we image?}
The spatio-angular imaging operator $\mc{H}$ only transmits six angular
components, so $\mc{H}$ can be decomposed into two operators:
$\mc{H} = \mc{H}'\mc{P}$, where
$\mc{P}: \mbb{L}_2(\mbb{R}^3\times\mbb{S}^2)\rightarrow
[\mbb{L}_2(\mbb{R}^3)]^6$ is a projection operator onto the direct sum of six
$\mbb{L}_2(\mbb{R}^3)$ spaces, and
$\mc{H}':[\mbb{L}_2(\mbb{R}^3)]^6\rightarrow \mbb{L}_2(\mbb{R}^2)$. The fact
that only six angular components are transmitted is a direct consequence of the
angular band limit imposed by dipole radiation.

However, angular structured illumination allows us to alias a much larger
number of angular components into the passband of the imaging system. For strong
excitation of dipoles undergoing slow angular diffusion, a theoretically
unlimited number of angular components can be aliased into the passband---see
Eq. \eqref{eq:angularfree} and Fig. \ref{fig:saturate}. Of course, the number of
angular components is practically limited by diffusion, photobleaching, and
noise---see Gustafsson \cite{gustafsson2005} for an analogous discussion of how
these factors affect spatial non-linear structured illumination. More
practically, weak excitation of dipoles undergoing slow angular diffusion allows
us to alias a total of fifteen angular components into the passband of the
imaging system (corresponding to the $\ell = 0,2,4$ spherical harmonics). For
fast-diffusing dipoles, aliasing does not occur and only six angular components
(corresponding to the $\ell = 0,2$ spherical harmonics) can be imaged.

Many other non-linear techniques can be used to alias high-frequency angular
components into the passband. Two-photon excitation beams excite with a
$\cos^4\theta$ dependence, so a two-photon beam can alias higher angular
frequencies than an equivalent single-photon beam \cite{ferrand2014}. A wide
variety of other techniques that exploit three- or multi-state fluorescent
molecules to alias high spatial frequencies can be adapted to alias high angular
frequencies---see the supplement of \cite{li2015} for a summary of non-linear
spatial structured illumination techniques. For a particular example, Hafi et
al. adapted stimulated emission depletion microscopy (STED) to the angular case
in a technique they called excitation polarization angle narrowing (ExPAN)
\cite{hafi2014}. Although they claimed their technique provided improved spatial
resolution, this claim has been challenged \cite{frahm2015} and we view ExPAN as
a technique that provides improved angular resolution that can be used to infer
higher spatial resolution if the coupling between spatial and angular
information is known.

We have focused on cases where the exposure time is much longer than 
the diffusion and decay times---so-called \textit{steady-state} experiments---but exposure times comparable to diffusion and decay times---so-called \textit{time-resolved} experiments---are possible \cite{suhling2004, becker2005, bowman2019}. We can model these experiments within our framework by changing the limits of integration in Eqs. \eqref{eq:intden} or \eqref{eq:intden2}. Time-resolved experiments allow for the measurement of more angular components than steady-state experiments, and we view these techniques as important future directions. 

\subsection{How many angular components can we reconstruct?}
Imaging an angular component is only the first step towards estimating an
angular component. To estimate a parameter it must be a linear combination of eigenfunctions of $\mc{H}^\dagger\mc{H}$ with non-zero eigenvalues (equivalently, a linear combination of the right singular vectors of $\mc{H}$ with non-zero singular values) \cite[Ch. 13.3]{barrett2004}. We will explore these functions---the so-called \textit{measurement space} of $\mc{H}$---in the next paper of this series.

For now we briefly mention two strategies for reconstructing multiple angular
components. The first approach is to take $N$ sequential measurements of the
same object after changing the illumination or detection polarization, then use
these measurements to reconstruct the angular components at each spatial
position independently \cite{axelrod1979, dale1999, siegel2003, demay2011a,
  mattheyses2010, brasselet2011, zhanghao2019}. This approach amounts to
approximating the complete imaging process
$\mc{H}_{\text{multi}}:\mbb{L}_2(\mbb{R}^3\times\mbb{S}^2)\rightarrow
[\mbb{L}_2(\mbb{R}^2)]^N$ with an imaging operator for each spatial point
$\mc{H}_{\text{point}}:\mbb{L}_2(\mbb{S}^2)\rightarrow \mbb{R}^N$. Although this
approach simplifies the reconstruction problem, it ignores valuable information
that can be exploited. We advocate for joint spatio-angular reconstructions that
use everything we know about the physics of the imaging process.

The second approach is to image single molecules and use their images to
estimate angular components \cite{backer2014, backer2015, zhang2018a,
  backer2019}. In this case the imaging process can be modeled with a single
imaging operator for each molecule
$\mc{H}_{\text{single}}:\mbb{L}_2(\mbb{S}^2)\rightarrow \mbb{L}_2(\mbb{R}^2)$.
This work's potential contributions to single-molecule imaging are improved imaging operators $\mc{H}_{\text{single}}$ that can be used to access more parameters (multiple diffusion constants or high angular frequency components).  

\subsection{Stochastic spatio-angular imaging}
A major limitation of this work is that we have only modeled the ensemble
average behavior of dipole diffusion, emission, and imaging, when these
processes are actually stochastic processes. More specifically,
angular diffusion is a random walk on the sphere, emission is an exponential
process, and photon imaging is a Poisson process. Existing works have modeled
these stochastic processes and then extracted the ensemble averages
\cite{lew2013, backer2015, stallinga2015}, while here we have modeled the
ensemble averages directly.

In this work we have focused on describing ensemble average features that have either been previously assumed absent (like the multi-exponential nature of diffusion) or not previously described (like non-linear angular structured illumination). Ultimately, optimal reconstructions will require complete stochastic descriptions of the imaging process so that correlations in the data can be exploited, and we consider stochastic descriptions of dipole imaging important future work.

\section*{Funding}
National Institutes of Health (NIH) (R01GM114274, R01EB026300, R35GM131843).

\section*{Acknowledgments}
TC was supported by a University of Chicago Biological Sciences Division Graduate Fellowship, and PL was supported by a Marine Biological Laboratory Whitman Center Fellowship. Support for this work was provided by the Intramural Research Programs of the National Institute of Biomedical Imaging and Bioengineering. 

\section*{Disclosures}
The authors declare that there are no conflicts of interest related to this article.

% \bibliography{/Users/Talon/Library/texmf/talon}

\appendix

\end{document}